# An electronic structure investigation of PEDOT with $AlCl_4^-$ anions, a promising redox combination for energy storage applications

Ben Craig[a], Peter Townsend[b], Chris Kriton-Skylaris[c], Carlos Ponce de Leon[a], Denis Kramer[a, d]

[a] School of Engineering, University of Southampton, University Road, Southampton, SO17 1BJ, UK

[b] Guest, Department of Chemistry and Chemical Biology, Rutgers University, 123 Bevier Rd, Piscataway, NJ 08854, United States

[c] School of Chemistry, University of Southampton, University Road, Southampton, SO17 1BJ, UK

[d] Faculty of Mechanical Engineering, Helmut-Schmidt-University, Holstenhofweg 85, 22043 Hamburg, Germany

## Abstract

The conducting polymer poly(3,4-ethylenedioxythiophene) (PEDOT) is one of the most researched materials. The 1980s bipolaron model remains the dominant interpretation of the electronic structure of PEDOT. Recent theoretical studies have provided updated definitions of key concepts such as bipolarons or polaron pairs, but these have not yet become widely known. In this work, we use density functional theory to investigate the electronic structure of PEDOT oligomers with co-located $AlCl_4^-$ anions, a promising combination for energy storage. By considering the influence of oligomer length, oxidation or anion concentration and spin state, we find no evidence for self-localisation of positive charges in PEDOT as predicted by the bipolaron model at the hybrid functional level. Our results show distortions that exhibit either a single or a double peak in bond length alternations and charge density. Either can occur at different oxidation or anion concentrations. We note that other distortion shapes are also possible. Rather than representing bipolarons or polaron pairs in the original model, these are electron distributions driven by a range of factors. Localisation of distortions occurs with anions, and distortions can span an arbitrary number of nearby anions. Conductivity in conducting polymers has been observed to reduce at anion concentrations above 0.5. We show at high anion concentrations, the energy of the localised, non-bonding anionic orbitals approaches that of the system HOMO due to Coulombic repulsion between anions. We hypothesize that with nucleic motion in the macropolymer, these orbitals will interfere with the hopping of charge carriers between sites of similar energy, lowering conductivity.

## Introduction

Despite having been used in a variety of applications, fundamental understanding of conducting polymers (CPs) is still incomplete and remains a highly active area of research. CPs are semiconducting in their pure form, and only become conductive when an oxidising or reducing species co-locates with the polymer chain via a redox reaction, a process conventionally referred to as doping by analogy with doped inorganic semiconductors. CPs have a carbon backbone consisting of alternating single-double carbon bonds [1, 2]. PEDOT continues to be the subject of particularly intensive research due to its chemical stability and tuneable conductivity [3]. To date, PEDOT has found applications in supercapacitors [4], sensors [5], solar cells, bioelectronics, electrochemical transistors, sensors, electrochromic displays [6], battery electrodes [7] and spintronics [8]. The present work studies the PEDOT/$AlCl_4^-$ system, inspired by the use of PEDOT as a cathode in a battery with $AlCl_3$:EMIm[Cl] electrolyte and an aluminium anode [9]. However, the co-locating of PEDOT with a small symmetrical anion [9] also allows comprehensive analysis of the electronic structure of PEDOT relevant to a wider field and is a useful complement to the more commonly studied, larger and less mobile anions such as polystyrene sulfonate (PSS) or tosylate (Tos) [1].

While many computational studies on PEDOT exist to date [10-14], to the best of our knowledge the present study is the first computational study that concerns the introduction of $AlCl_4^-$ onto PEDOT, except the authors' own prior work [15, 16].

### *Structure and electronic properties of PEDOT, and the bipolaron model*

In conventional semiconductors, the high coordination number (four and greater) of covalent bonds between atoms to their neighbours provides a rigid structure such that electronic excitations can be viewed as electrons or holes in a lattice. In CPs, each monomer unit is bound to a maximum of two other monomer units by carbon-carbon bonds, making CPs subject to structural distortions caused by the addition or removal of electrons. Much of the early theoretical work on conducting polymers focussed on attempting to fit existing semiconductor theory to these new materials.

It has been consistently observed and predicted that when counterions are added to conducting polymers, localized charge-carrying distortions occur over a length of a few monomer units [17, 18] as shown in Figure 1. A key feature is that the change in bond lengths observed closely corresponds to the charge density. Electron kinetic energy drives increased distortion length, while the lattice relaxation drives decreased distortion length; the resulting length is a balance of both [14, 19]. The theory of few-monomer distortion lengths is based on the observation that the distortions in conducting polymers are localized around the part of the chain where the anions are located, and their orbitals do not mix enough to form extended energy bands [20]. Instead, they form a series of localized, overlapping states with significant electronic and structural distortions [21]. The fact that the distortions are localized has led researchers to term them polarons.

The bipolaron model in its original form was developed from theoretical and experimental results in the 1980s, shortly after conducting polymers were discovered, in an attempt to explain some of their key characteristics. For example, in inorganic semiconductors, trapped polarons would be expected to build up throughout the lattice as they are oxidised. In contrast, when a conducting polymer is oxidised, the spins build up, before disappearing again, with no spins detected in the fully charged polymer [18, 22, 23]. To explain this, the bipolaron model was devised, which stated that spins first appear as localised singly charged states termed polarons with spin ½, and then as the polymer is further oxidised, the charges co-locate into doubly-charged spinless states termed bipolarons, explaining the disappearance of the spins [18]. Polarons and bipolarons according to the original 1980s bipolaron model are shown in Figure 1.

The single-double bond pattern in the conjugated backbone is critical to the characteristics of conducting polymers. For the simplest conducting polymer, trans-polyacetylene, the hydrogen atoms are on alternating sides of the carbon atoms. The conjugated carbon backbone has alternating single-double bonds due to a Peierls distortion [24]. No change of energy results from inverting the single-double bond pattern: that is to say, the two possible orientations are degenerate. The presence of a counterion introduces a local region with inverted single-double bond patterns, termed a soliton in the 1980s bipolaron model [25].

Thiophene derivatives including PEDOT have three carbon-carbon bonds in each monomer unit, and so bond inversion is not energetically neutral. A neutral PEDOT monomer unit has a long central bond and a shorter bond either side, described as aromatic structure. The inverted pattern appears in the presence of anions; this is described as quinoid [25, 26] as shown in Figure 1. A polaron is the term given to a localized single-charged distortion, with an accompanying surrounding lattice distortion carrying a distinct charge, analogous to polarons found in inorganic polar insulators and semiconductors [25]. PEDOT and polythiophene are typically treated with anions, becoming oxidised (p-doped), so the polarons are hole polarons [27].

When the polymer is further oxidised, so that it carries two charges per oligomer, the bipolaron model predicts that the charges locate together into a doubly charged excitation termed a bipolaron [28], illustrated in Figure 1. The theoretical driving force is that the energy saved by structural relaxation is greater than the energy penalty from Coulombic repulsion between the two like charges [14]. Further increases in anion concentration lead to multiple bipolaron formation [18]. The 1980s bipolaron model was supported by predictions from early theoretical methods such as one-electron extended Hückel theory. For example, Hückel theory finds that the energy of distortion to form two polarons or one bipolaron is quite similar, but the ionization energy is less for a bipolaron than two polarons [18]. Hückel theory predicted polarons and bipolarons to be around 4 monomer units long in polypyrrole [29]. A semiempirical analysis using the modified neglect of diatomic overlap (MNDO) method showed similar results [30].

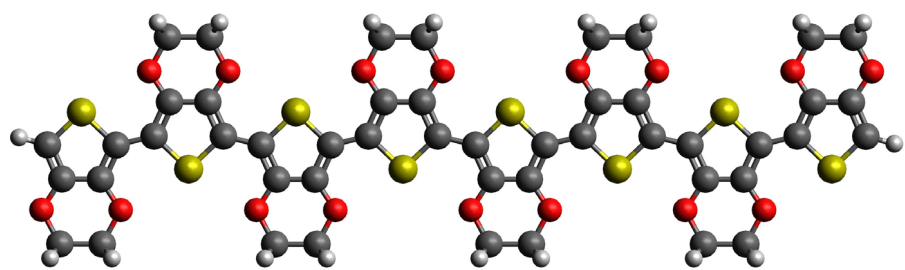
Neutral – aromatic structure throughout

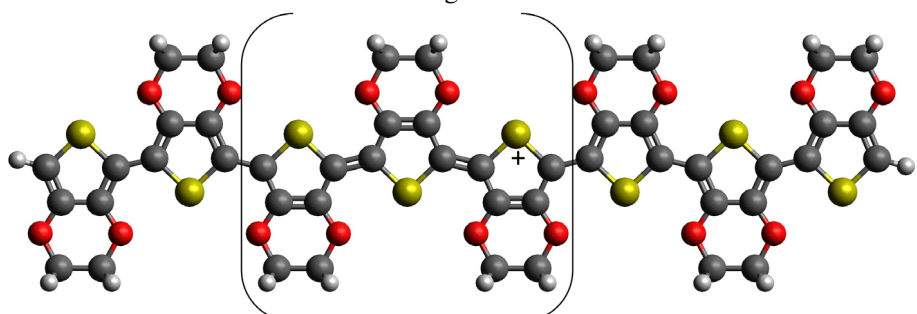
Polaron, one electron removed; quinoid structure between brackets

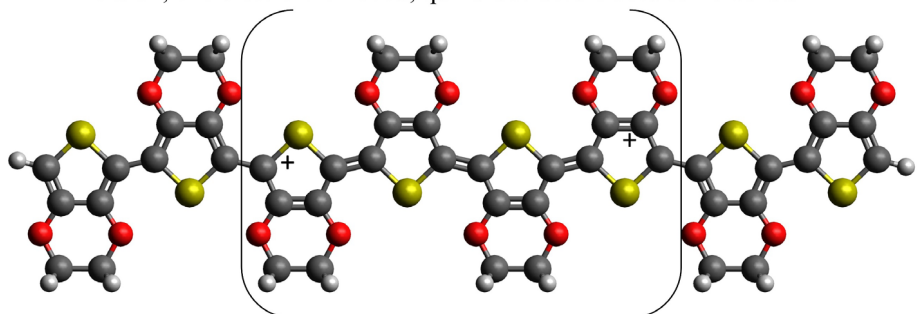
Bipolaron, two electrons removed; quinoid structure between brackets

**Figure 1 – 8-PEDOT according to the original 1980s bipolaron model, demonstrating the proposed form of a polaron and bipolaron compared to a neutral oligomer, with accompanying inversions of the single-double carbon-carbon bonds. Yellow atoms are sulfur, red atoms are oxygen, grey are carbon and white are hydrogen. The positive charges are indicated on single carbon atoms with three single bonds, but are distributed over a few monomer units as indicated by the brackets. The bond lengths are not to scale and are illustrative only. Adapted from the review by Heinze et al. [17]**

When conducting polymers are in their pure state and unoxidised, they have a bandgap characteristic of semiconductors and the HOMO-LUMO gap is too large to allow much or any conduction. The bipolaron model, set out by Brédas *et al.* [18] and based on an infinite periodic polymer with a band structure, predicts that the removal of an electron from a neutral conducting polymer causes the half-filled energy level to be pushed upwards into the gap. The half-filled level is termed the polaron level. An empty state also appears below the conduction band, termed the bipolaron level. The appearance of these states reduces the bandgap. Removing a further electron causes the newly unoccupied polaron level to be pushed up further still, with a similar decrease in energy of the bipolaron level, further reducing the bandgap [18].

Brédas *et al.* also predicted that the valence band, not including the half-filled level which is described as the polaron band under the bipolaron model, would always be full [18]. It was proposed that the levels appearing in the bandgap corresponded to the π (bonding) and π* (antibonding) states created by the mixing of the anionic states with the existing system

[29]. Regardless of debate over the bipolaron model, it is well known that the reduction of the bandgap via the introduction of anions is a key contributor to increased conductivity.

The actual mechanism of conductivity in conducting polymers is a mix of intra- and interchain electron and hole mobility. Conductivity along a PEDOT molecule occurs once there are sufficient numbers of well-distributed anions adsorbed onto it. This causes the quinoid regions to overlap and pass electrons or holes down the conjugated carbon backbone via resonance between many states of nearly identical energy, with many orders of magnitude larger conductivity than the pure polymer [21]. This is possible via rapid swapping of the single-double bonds which allow the charge carriers to travel efficiently [5]. This mechanism means that conductivity along the PEDOT molecules is typically very high at moderate anion concentrations. This movement can be thought of as intrachain hopping of charge carriers between nearby sites of similar energy [4].

Hopping or tunnelling of charge carriers between different PEDOT molecules and regions of the polymer is also required for conductivity. PEDOT and other conducting polymers are generally a mix of ordered and disordered regions with some degree of self-ordering due to π-π interactions. In the ordered regions, the conductivity depends on the direction of hopping in the stacked polymer. Various models have been developed or adapted for interchain conductivity [4], including Mott's Variable Range Hopping (VRH) [31] or Sheng's fluctuation induced tunnelling [32]. Further work has suggested that only some regions in the PEDOT achieve polaronic structure and that hopping must also take place between these highly conductive regions [33]. In general, interchain hopping between different molecules or regions has been found to be the limiting factor in conductivity [4].

The bipolaron model predicted that as the concentration of counterions rises, the spinless bipolarons become mobile, and are able to conduct due to the broadening of the bipolaron band, which forms from the bipolaron level as many electrons are removed [17]. At the highest counterion concentrations, the bands can even merge to produce a metal or semimetal [21]. This explanation fits well with the experimentally observed increase in conductivity of conducting polymers at low anion concentrations.

However, experiments on conducting polymers including several thiophene derivatives, albeit not including PEDOT, have shown that conductivity increases up to counterion concentrations of 0.5, then decreases above this [34], for which the bipolaron model offers no explanation [17]. The same behaviour was observed for a hexathiophene and polyphenylene hybrid [35]. Heinze *et al.* suggested that conducting polymers have a maximum of conductivity at 0.5 due to the same mechanism as radical ion salts [17, 36]. This is proposed to be the point at which there is a maximum of charge carriers available to hop between sites of similar energy on different chains. Heinze *et al.* proposed that during the wide plateau of conductivity, overlapping redox states are reached successively, keeping the system in a mixed valence state until the last redox sites are emptied and the conductivity drops [17]. Meanwhile, Brédas *et al.* attributed this discrepancy with the bipolaron model to screening of the attraction between counterions and their polarons at high anion concentrations [18]. Others have found that at high anion concentrations, the anions can cluster into regions which inhibit electron hopping between PEDOT regions [37]. Ofer *et al.* suggested that the presence of both charged and uncharged sites would be required for charge carrier hopping. They theorized that the loss of conductivity at high anionic concentrations was due to some kind of localization of charges due to Coulombic repulsion between solvent and charges in the highly charged states [34].

The research in this area is limited partly because of a generally accepted view that anion concentrations normally reach a maximum of about 0.4 in electrochemically synthesized conducting polymer films [17, 38, 39]. Higher levels are achievable, but special electrolytes are required to reach concentrations above 0.5 [34]. No studies could be found that examined PEDOT conductivity specifically as a function of calculated anion concentrations above 0.5, but several show a trend for conductivity to drop as the concentration of anions available for adsorption increases past a certain point [40-43].

What all these mechanisms have in common is that they rely on charge carriers hopping between different sites of similar energy. In this work, it is found that at high anion concentrations of 0.66 on single PEDOT molecules, the Coulombic repulsion between the anions pushes up the energy of their highly localised orbitals so that they are close in energy to the HOMO of the overall system, which is normally a delocalised orbital extending over the conjugated carbon backbone. In this thesis, it is suggested that combined with movements of the polymer at temperatures above 0 K, these highly localised orbitals might be pushed up in energy sufficiently that they begin to interfere with the hopping mechanism between different sites of similar energies, affecting both intra- and interchain conductivity.

### *Progress from density functional theory*

Since the 1980s when the original bipolaron model was developed, DFT, particularly hybrid DFT incorporating a proportion of Hartree-Fock exact exchange energy, has become the de facto standard approach for modelling conducting polymer systems due to its ability to handle electron correlation effects while remaining efficient for reasonably large systems [16]. Investigations via DFT have provided significantly different insights into the electronic structure of conducting polymers and have led to revised explanations of experimental observations.

A key limitation of one-electron theories is that they omit electron-electron repulsion [14]. One-electron methods assume degeneracy of the spin orbitals even for spin-asymmetric configurations with a partially occupied level, treating the half-filled band as fully occupied. The spin degeneracy assumption distorts the results between odd and even oxidation levels, as the independence of spin-up ($\alpha$) and spin-down ($\beta$) orbitals is an important part of the energy minimization process for any open-shell system. Consequently, Hückel theory predicts that two new spin-degenerate levels are created in the gap for each oxidation step of the PEDOT; a lower singly occupied state and upper unoccupied state for a polaron, and two empty states for a bipolaron [44]. In contrast, *ab initio* methods do not predict spin degeneracy for polarons, and only one state appears in the gap [6]. Meanwhile, semiempirical methods need parameterising carefully from appropriate *ab initio* methods [14].

Hybrid DFT calculations generally do not support the existence of the type of spinless bipolaron predicted by the one-electron or semiempirical approaches where charges co-locate by energetic preference of structural relaxation [6, 45]. A review of DFT studies indicates that bipolarons do not exist in long oligomers without counterions [14]. Only oligomers shorter than eight units long [14], and oligomers with one counterion per six or fewer thiophene rings [46] exhibit bipolaron-like behaviour [10], and then only because there is not enough room for the distortions to separate into polarons. Also, spectroscopic experiments on oligothiophenes have demonstrated that the sub-gap transitions previously attributed to bipolarons were actually two separated polarons in long oligothiophenes [47]. Findings such as this led to the invention of the term polaron pair, which consists of two polarons for which it is not energetically favourable to co-locate, so they tend to move apart along the chain if there is space to do so. In this conception, bipolarons are closed-shell singlet states and therefore spinless, while two polarons on the same chain can be singlet or triplet biradicals [48].

The presence of counterions also serves to pin charge carriers in a particular location, which can give rise to apparent bipolaronic behaviour, generating overlapping distortions that appear as one larger one [25]. Modelling of thiophene chains of various lengths up to 25 monomer units has indicated that polarons tend to spread out but do interact still in the middle of the chain, observed via bond length changes compared to neutral oligomers, implying an interaction spread over a much larger part of the chain. These results have also been interpreted to support the concept of polaron pairs rather than bipolarons, referring to the triplet state reported for oligomers longer than ten monomer units with +2 oxidation state which shows two peaks in bond length alternations [49].

In contrast to previous explanations of the build-up and disappearance of spins, more recent DFT work has proposed that the tendency for spins to build up during oxidation and then reduce or disappear once oxidation is complete is caused by the removal of multiple single electrons from fully occupied valence states across different chains in the polymer, and then the subsequent removal of the singly occupied electrons, leaving the original HOMO-1 as the new, fully occupied HOMO [6]. The differences between the bipolaron model and the modern view are subtle, but they can be summarised as that the bipolaron model predicts the incorrect band structure and predicts that two like charges have a structural preference for co-location in oxidised conducting polymers with and without counterions. Modern DFT research unanimously rejects the band structure of the original bipolaron model and generally rejects the idea of a structural tendency for two like charges to co-locate in oxidized conducting polymers without anions, at least not at the ground state, although at higher multiplicities single, larger distortions can appear. The presence of nearby counterions in oxidised conducting polymers that are relatively fixed in position can also generate larger, combined distortions due to overlapping distortions.

To account for the developments from modern theoretical methods, it has become convention to refer to the original 1980s bipolaron model as the 'traditional model' or its predictions as those of the 'traditional methods'. Since 2019, some authors have adapted the terms from the original bipolaron model to fit modern understanding [6, 50]. Perhaps the clearest elucidation of the updated terms polaron, bipolaron and polaron pair is provided by Sahalianov *et al.*, with a polaron referring to a doublet with +1 charge, a polaron pair to a triplet with +2 charge, and a bipolaron to a singlet with +2 charge [50]. In this approach, states with more than two charges are referred to as polaron(ic) or bipolaron(ic) for odd and even charged states respectively [6, 50]. An important feature is that the terms are no longer linked explicitly to distortion characteristics, unlike less recent theoretical work [10, 48, 49]. Sahalianov *et al.* also updates the definition of a bipolaron to mean a distortion shared between two nearby anions, and a polaron pair to mean two anions that are far enough apart to produce two separate distortions, but with no structural preference for co-location of charges implied for either [50]. Ambiguity remains over how to describe what happens when more than two anions are near enough to share a single distortion.

This change in definition means that a bipolaron or a polaron pair is no longer associated to a distortion shape or to an energetic preference for co-location or separation of two charges, leaving very little of the original meaning. It is argued that this adaptation of the original language is confusing and presents a barrier to accessing modern theoretical understanding for experimentalists and those new to the field. This is evidenced by the fact that many researchers are still using the bipolaron model in its original 1980s form based on pre-DFT methods [17, 21, 27, 39, 51-58] as also pointed out elsewhere [6, 50]. The picture from modern theoretical methods is not confusing when explained from first principles, as is demonstrated in this work. For clarity and because others still base their understanding on it, we adopt the convention of referring to the bipolaron model in its original 1980s form.

It is important to note that the investigation of oxidised conducting polymers without counterions is really a hypothetical study problem. In the battery context, and in other real-life contexts, it cannot be expected that pure PEDOT can be oxidised without counterions present to balance the charges. Despite this, conducting polymers without anions have been thoroughly investigated by now. Therefore, a more pertinent research question going forwards is to focus on the electronic structure of conducting polymers with counterions.

## Results and discussion

## Functional choice for modelling conducting polymers

Historically, B3LYP has been used extensively for modelling conducting polymers [10, 11, 14, 48, 59-62]. However, recently, there has been some movement towards the range-separated hybrid function wB97XD [6, 12]. The main reason for moving away from B3LYP is due to its over-delocalization of electrons, with no localisation of distortions found in singly charged oligomers without anions. However, much of the research has focussed on polymers without anions. This is an unphysical situation as the polymers cannot be oxidised without counterions in experiment. The introduction of anions is known to pin distortions locally. Therefore, it was important to test whether wB97XD predicts different results from B3LYP for a number of test cases: a neutral 12-PEDOT without anions, 12-PEDOT at +1 and +2 oxidation states without anions, and 12-PEDOT with one and two anions.

Figure 2 shows the neutral 12-PEDOT oligomer and the 12-PEDOT +1 doublet without anions. The neutral case shows almost identical results in terms of charge distribution. In terms of quinoid *vs*. aromatic regions, where a monomer unit has aromatic structure if the bond lengths form an arrow shape pointing upwards and quinoid if the arrow shape points downwards, all monomer units are aromatic for both functionals. However, wB97XD predicts a greater difference between short and long bonds. For the neutral case then, there is qualitatively very little difference between the two methods.

For the 12-PEDOT +1 doublet without anions, there is a pronounced difference between the two functionals. The bond lengths of wB97XD shows a quinoid region localised over two monomer units with a sharp transition on the monomer units either side. In contrast, B3LYP predicts that the difference between short and long bond lengths slightly reduces. This is a key difference between the two functionals, with B3LYP tending towards a delocalisation of distortions along the whole oligomer in PEDOT without anions while wB97XD localises strongly. This has also been shown elsewhere as previously discussed [12]. The charges follow the distortion location: wB97XD consequently has a charge peak around the middle of the oligomer. The slight asymmetry in summed charges is attributed to a rounding error in the partial charge calculation.

The results for 12-PEDOT +2 singlet and triplet without anions are shown in Figure 3. The results for each functional are much more similar for the singlet case. wB97XD results in a distortion that is one monomer unit shorter in length at each end, showing a quinoid region spanning six monomer units rather than the eight for B3LYP. Again, the charge profile for B3LYP is highly delocalised. In contrast, the wB97XD has a charge profile that has a wide plateau over the length of the distortion. These findings confirm a significant difference between the two functionals for PEDOT without anions, and given the known over-delocalisation of B3LYP, suggests the wB97XD may offer more accurate insights.

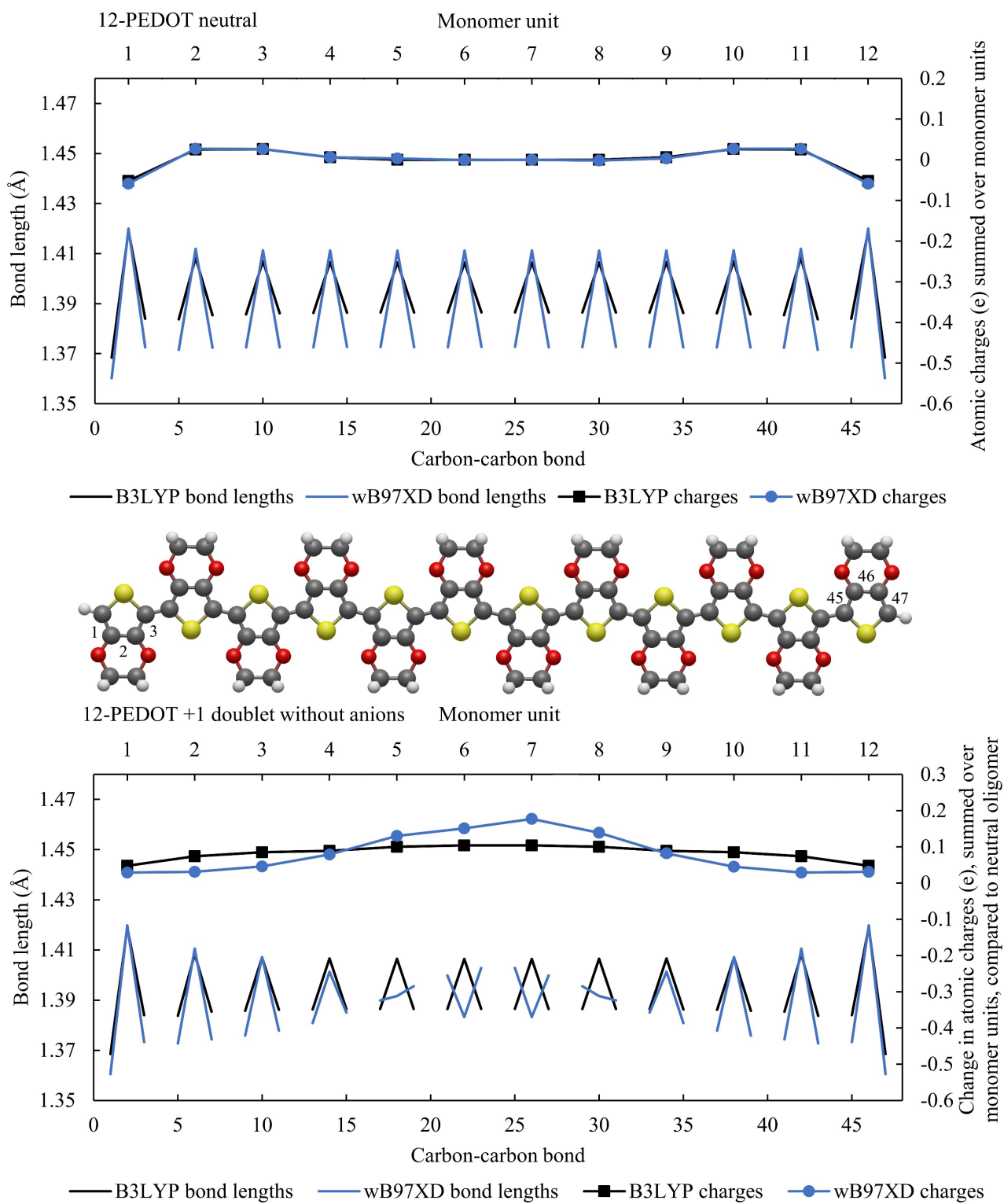

Figure 2 - 12-PEDOT modelled with B3LYP and wB97XD, showing bond lengths and summed charges on each monomer unit. For the neutral case (top), the charges are summed. For the 12-PEDOT +1 without anions, the charges are the difference in summed charges per monomer unit from the neutral oligomer. The carbon-carbon bonds between monomer units are negated in order to highlight the aromatic vs. quinoid structure.

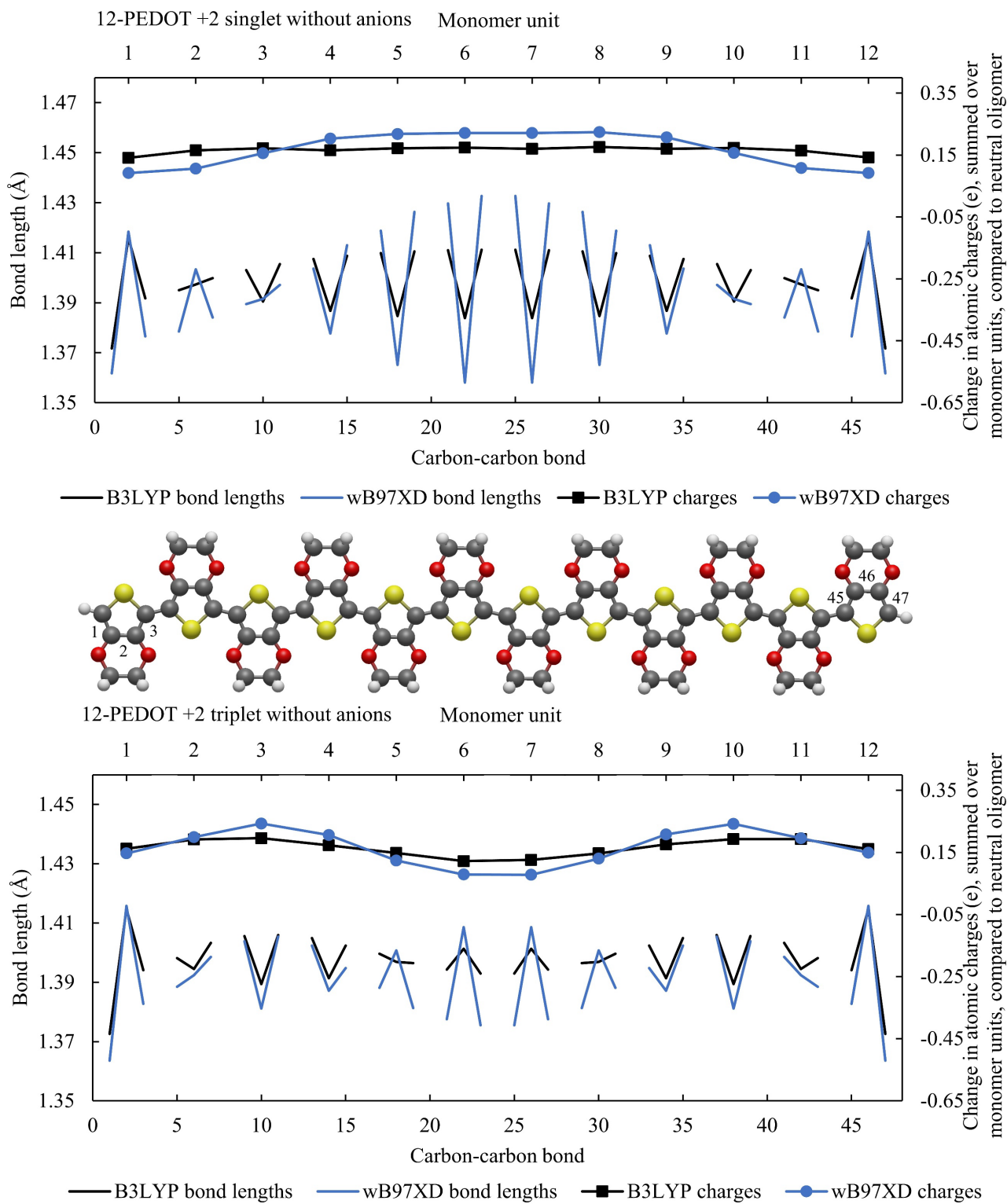

**Figure 3** - 12-PEDOT +2 singlet and triplet modelled with B3LYP and wB97XD, showing bond lengths and summed charges on each monomer unit. The charges are the difference in summed charges per monomer unit compared to the neutral oligomer. The carbon-carbon bonds between monomer units are negated in order to highlight the aromatic vs. quinoid structure.

To analyse the system with anions, 12-PEDOT with one anion (singlet ground state) and 12-PEDOT with two anions (doublet ground state) are shown in Figure 4. In both cases, the areas of quinoid and aromatic structure are much more similar between the two functionals than for oxidised PEDOT without anions, with the only difference qualitatively being that in the case of 12-PEDOT with two anions, wB97XD transitions from aromatic to quinoid between one monomer unit and the next, while B3LYP shows a transition monomer unit with structure that is not clearly aromatic or quinoid. With anions, the summed charges per monomer unit relative to the neutral oligomer are also much more closely matched between functionals. It is a known error of the correction for the self-interaction error that DFT systematically underestimates charge transfer in redox reactions when this is known in reality to be almost always complete. Therefore, the dip in charge around the anions can be attributed to negative charge spilling back from the anion. wB97XD predicts a smaller error of this kind, shown by a smaller dip in the summed charge, but the error persists. These results show that B3LYP does not predict significantly more delocalized distortions when anions are present compared to wB97XD, suggesting that either functional can be used to produce similar results on this system.

While this work focusses on PEDOT with $AlCl_4^-$ anions, it is worth noting a few key observations from these graphs. Both functionals show a single peak distortion in bond lengths and charge distributions for 12-PEDOT+2 singlet, and a double peak distortion for 12-PEDOT+2 triplet. Noting that the exact energy levels of the ground states are not very reliable from any DFT functional, it can be said that either singlet or triplet state could exist for the 12-PEDOT+2 without anions. Furthermore, the peaks in the charge distribution do not add up to a double or two single charges as would be expected for a bipolaron or two separated polarons (a polaron pair). The same observation can be made from Figure 2 for 12-PEDOT+1. Instead, the charge is distributed over the full length of the oligomer, even for wB97XD. This directly contradicts the original bipolaron model that a polaron or bipolaron are linked to particular distortion characteristics, and in this case, both single and double peak distortions are possible for the same oxidation state.

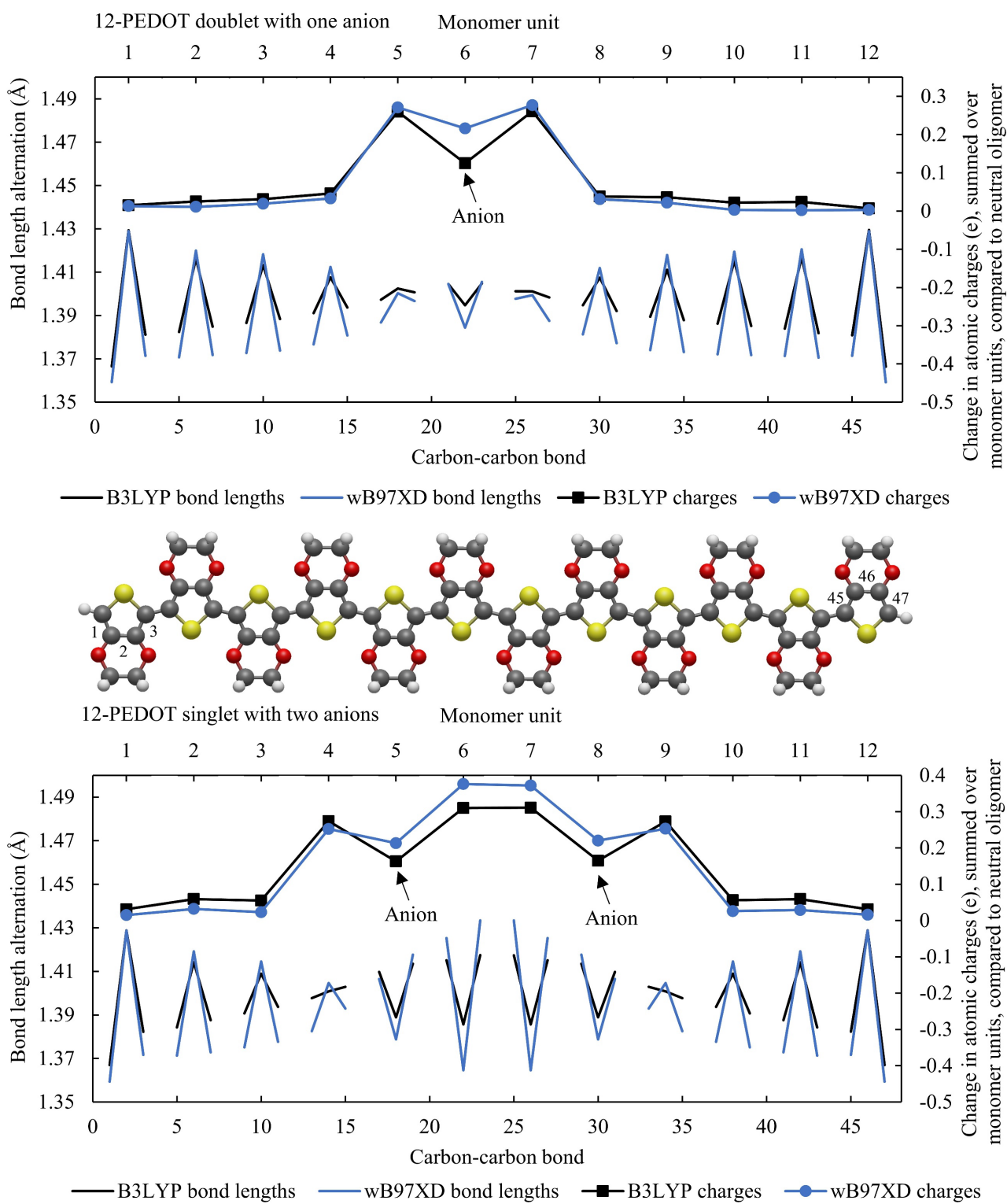

Figure 4 - 12-PEDOT with one and two anions modelled with B3LYP and wB97XD, showing bond lengths and summed charges on each monomer unit. The charges are the difference in summed charges per monomer unit compared to the neutral oligomer. The carbon-carbon bonds between monomer units are negated in order to highlight the aromatic vs. quinoid structure.

It is notable that delocalization error is one of the biggest remaining challenges to overcome with DFT and no existing functional offers a complete solution [63]. It is responsible for predicting incomplete charge transfer for redox pairs when unity charge transfer is expected as well as the aforementioned delocalization of electrons. Meanwhile, static correlation error is linked to inaccuracies in estimations of the energies of degenerate states, which can sometimes be avoided by using broken symmetry calculations. This problem is linked to the fact that DFT uses a single Slater determinant: it is monodeterminantal. Multideterminantal methods are used in situations where this is a problem, but they are not possible for larger systems [64]. During calculations for the current work, attempted broken symmetry calculations always failed to converge, and the issue is avoided by discussing only the qualitative electronic structure of the different low-lying excited states, rather than drawing firm conclusions about which is the true ground state.

In summary, while B3LYP and wB97XD predict quite different results for oxidised PEDOT without anions, the differences between them are small for neutral oligomers and when anions are introduced. Therefore, in the study of polymer systems with anions, it is not evident that range-separated hybrid functionals provide an advantage and therefore B3LYP is used throughout the rest of this study.

# Results and discussion

## Bond length and partial charge analysis

When $AlCl_4^-$ anions are introduced to the system, they locate in stable energy wells over the sulfur atoms and between the two oxygen atoms, out of the plane of the PEDOT oligomer, and are able to site either side of the plane of the fused rings. Previous work has shown that this position is generally stable providing only one anion is incorporated per PEDOT oligomer, though if several anions are positioned on adjacent PEDOT oligomers, they may spread out to adjacent sites during DFT structural relaxation or *ab initio* molecular dynamics [15]. This result contrasts with the original bipolaron model which predicts an energetic preference for co-location of charges.

To examine the effects of different numbers of anions on the chain, a 6-PEDOT and 12-PEDOT system were considered with a number of anions from zero to four, which could be placed anywhere except the end monomer units. Low energy configurations were chosen from the full range of possible configurations for 6-PEDOT and a large random sample of configurations for 12-PEDOT. The actual configurations used can be seen in the Supplementary Material, Figure 11.

Figure 5 shows the effect of introducing one or two anions to 12-PEDOT in terms of bond lengths and electron distribution. The configurations chosen are the lowest energy anion configurations from a large sample. When a single anion is introduced on monomer unit 6 (12-PEDOT +1), the bond length alternations localise over just the three monomer units nearest the anion, strongly localising the distortion. This result demonstrates that the introduction of anions results in highly localised distortions. For the +2 system with counterions, 12-PEDOT was set up with anions on monomer units 5 and 8. The charge profile and bond length alternations show a distortion six monomer units long, showing that where two anions are near together, a single distortion can occur spanning both.

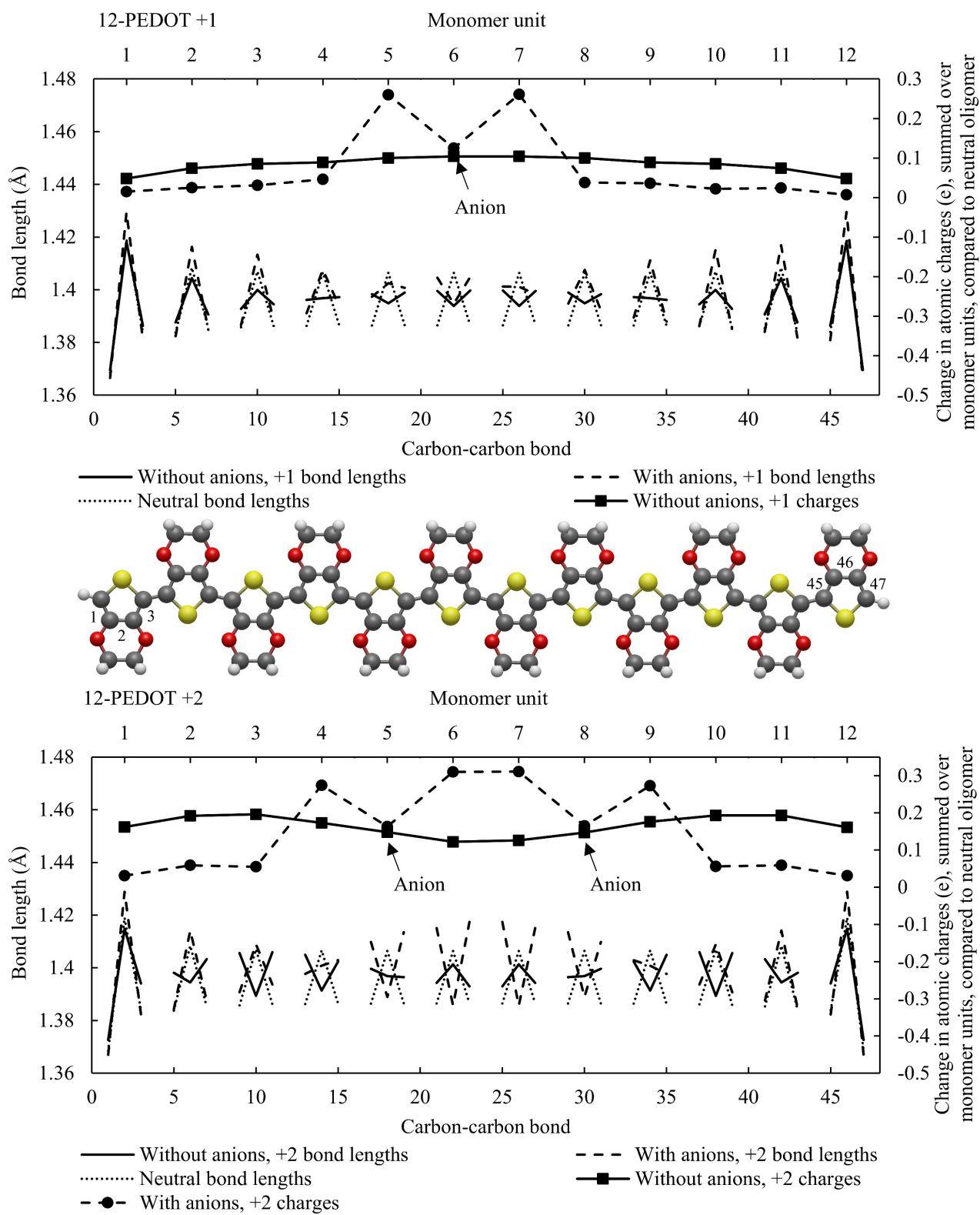

Figure 5 – Changes in atomic partial charges calculated from the ESP summed over monomer units compared to the neutral oligomer, and bond lengths of the carbon-carbon bonds. Upper panel: changes in charge density compared to the neutral oligomer, and bond lengths for 12-PEDOT, both without counterions at +1 oxidation state, and with a single anion on monomer unit 6. The bond lengths of the neutral oligomer are shown for reference. The end three units remain sharply aromatic in both cases as previously observed, indicating a continued tendency for negative charge to collect at the ends of the chain. The system has a doublet ground state. The anion causes the bond length alternation (and area of positive charge) to locate on a smaller region only three monomer units long, demonstrating how anions pin the distortions in place. Lower panel: 12-PEDOT +2 without counterions shows the triplet (ground) state, and with anions,

the singlet (ground) state with two anions on monomer units 5 and 8. The end units remain sharply aromatic in both cases, indicating a continued tendency for negative charge to collect at the ends of the chain. The 12-PEDOT without counterions, which has a triplet ground state, demonstrates two separate distortions and charge peaks while the system with anions, which has singlet ground state, forces the excitation to locate on the four monomer units closest to the anions, with a peak of positive charge in this region. As observed for 12-PEDOT without anions in Figure 3, there are two general forms of distortion that appear in the PEDOT systems studied in this work: a single or double peak distortion. It is found here that both can occur for PEDOT systems with anions as well. Figure 5 shows that for 12-PEDOT with anions, a single peak distortion occurs for one and two anions, noting that the charge density exhibits a dip on the monomer unit where the anion is located, with no accompanying change in the bond lengths. This is attributed to DFT incorrectly calculating the anions as having less than -1e charge. This single peak distortion observed is a similar result to that presented by Sahalianov *et al.* who found that two counterions a relatively short distance apart would cause a single peak distortion with singlet ground state, which they referred to as a bipolaron, whereas two anions further apart could induce a double peak distortion with triplet ground state, which they termed a polaron pair [50]. The possibility of a polaron pair in this definition is not explored here: to observe it, Sahalianov *et al.* used an 18-oligomer, longer than is explored in this work.

Figure 6 shows that for 6- and 16-PEDOT, when 4 anions are placed on adjacent monomer units on the oligomer to examine the effects of a high anion concentration on a localised part of a longer oligomer, a strong double peak distortion occurs. This implies that higher oxidation states are more associated with double peak distortions. In the case of 16-PEDOT with 4 anions in Figure 6, the pinning of the distortion around the anions is seen as the distortion only extends over eight monomer units.

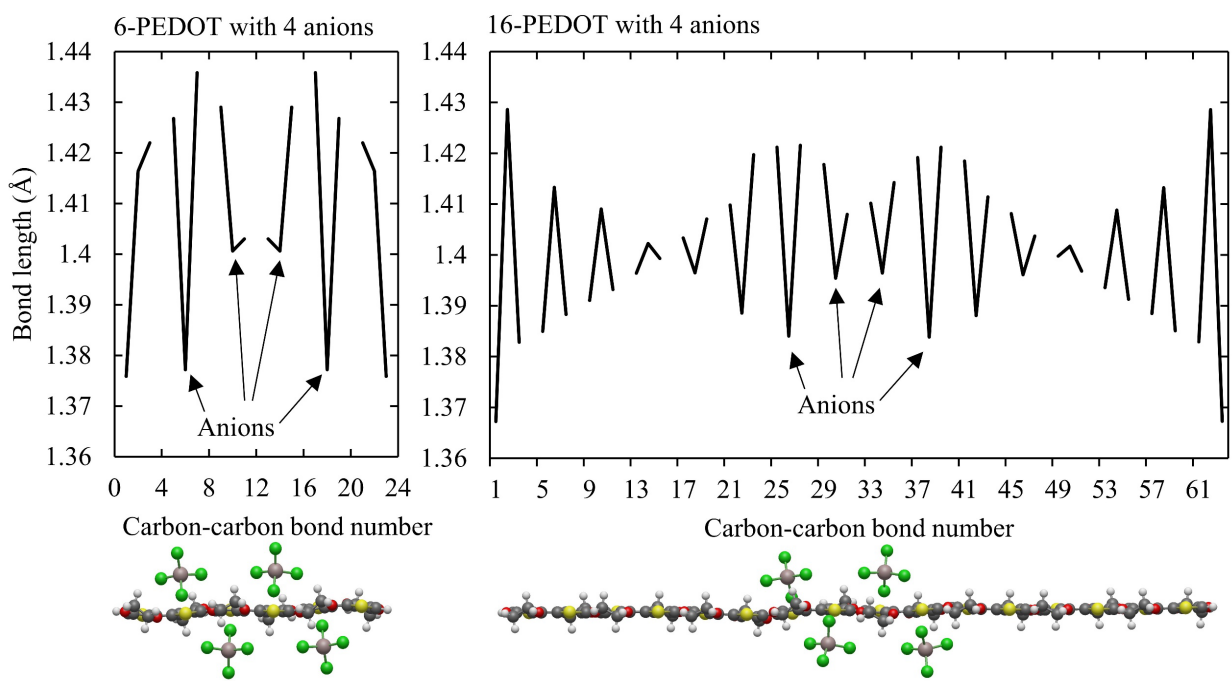

**Figure 6 – Bond lengths for 6- and 16-PEDOT with four anions. Left panel: bond lengths for 6-PEDOT with four anions on alternating sides of the central four monomer units, with the anion locations shown on the PEDOT oligomer below. A pronounced double peak region of quinoid structure is seen. Right panel: bond lengths for 16-PEDOT with four anions on alternating sides of the central four monomer units, with the anion locations shown on the PEDOT oligomer below. A pronounced double peak region of quinoid structure is also seen, though it is localised towards the middle of the**

oligomer due to the location of the anions. ## Energy level structure of neutral and oxidised 6- and 12-PEDOT with anions

Figure 8 plots the energy levels for these systems for an energy range around the HOMO-LUMO gap with zero to four anions. For comparison, Figure 7 presents the same energy levels for neutral and oxidised PEDOT without anions. The HOMO-LUMO gap gets progressively smaller as anions are added.

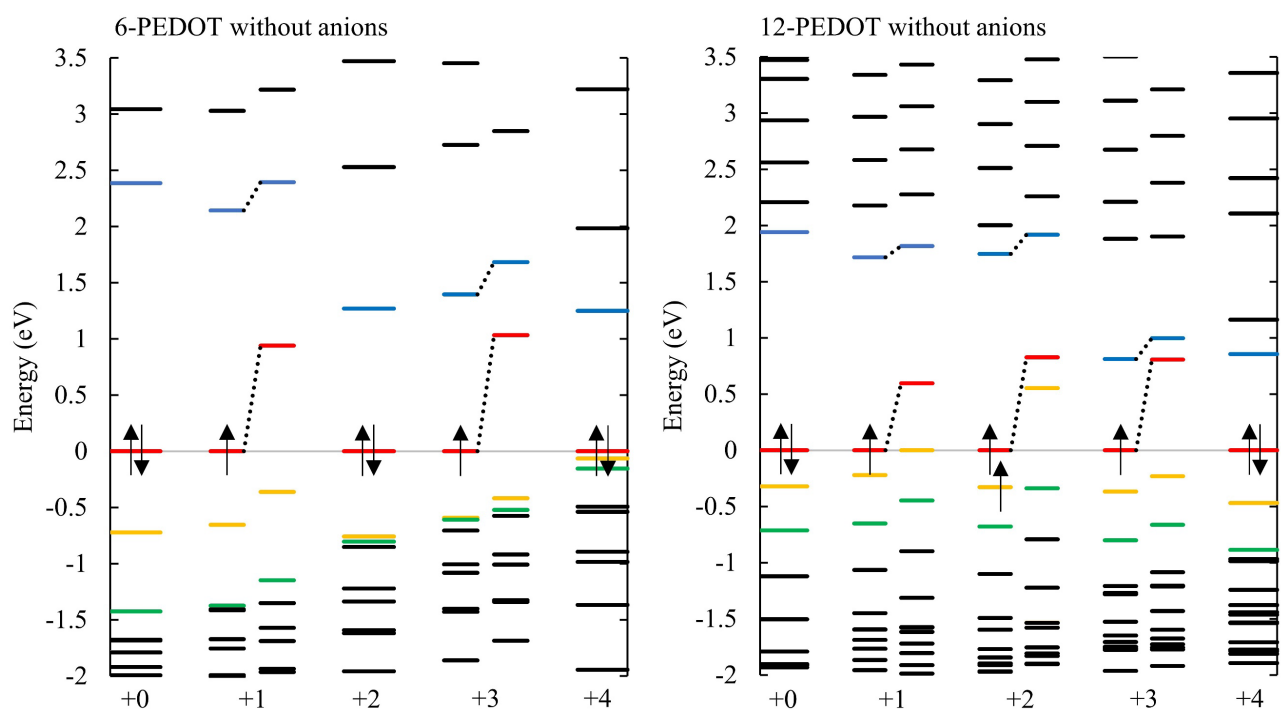

Figure 7 - Energy levels for 6-PEDOT and 12-PEDOT, without counterions, at different oxidation states. For filled valence shells the HOMO is indicated with both up and down spin electron arrows, while for partially filled valence shells (including the triplet) only the partially filled α levels are indicated. It is shown that for odd charges, with doublet ground state, a single unoccupied layer appears in the gap, which is the unoccupied β orbital of the HOMO. For 12-PEDOT, the patterns are similar, but the HOMO-LUMO gap reduces faster for longer chain lengths. For the +2 state, which has triplet ground state, two unoccupied levels appear in the gap; these are the unoccupied β orbitals of the two singly occupied orbitals. The HOMOs are normalized to 0 eV. Without this normalization, the overall energy level scale reduces as

**electrons are removed.**

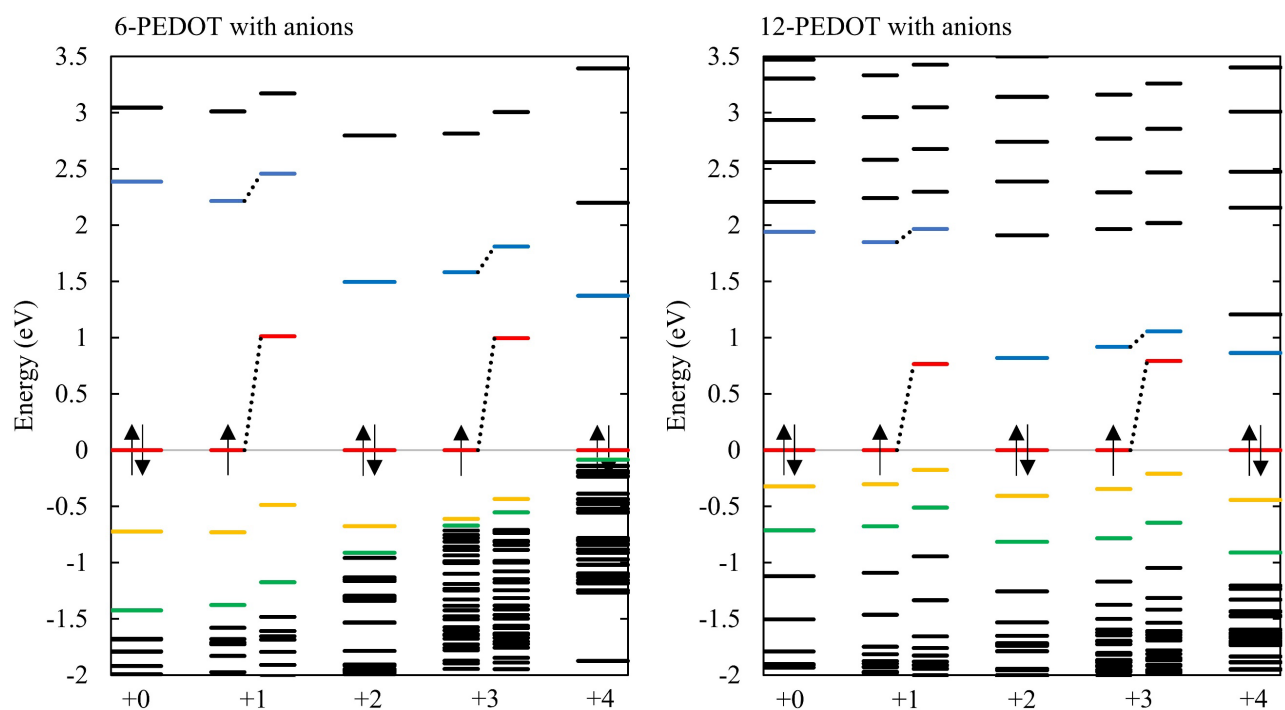

Figure 8 – Energy levels for 6- and 12-PEDOT with different numbers of anions. For filled valence shells the HOMO is indicated with both up and down spin electron arrows, while for partially filled valence shells only the partially filled α HOMO are indicated. At high doping levels for 6-PEDOT there is a close grouping of energy levels below the HOMO because the non-bonding orbitals of the anions have been introduced. These are heavily localised on the chlorine atoms of the anions (see Figure 9) and so can bunch closely as they do not interact with each other. Otherwise, the plots with anions are qualitatively similar to those without counterions (Figure 7) with a general decrease in bandgap as electrons are progressively removed from the PEDOT oligomer. For 12-PEDOT compared to 6-PEDOT, the close clusters of energy levels occurs further down the energy scale because for 12-PEDOT, the overall anion concentration is lower and the anions are more spread out. The triplet ground state seen in Figure 7 for 12-PEDOT disappears for 12-PEDOT with two anions, which has singlet and doublet ground states throughout. The configurations are shown in Supplementary Material Figure 11. The HOMOs are normalized to 0 eV as for Figure 7.The energy level diagrams corresponding to the PEDOT without and with anions in Figure 7 and Figure 8 respectively are very similar in character, showing that the oxidation level of the PEDOT rather than the position or presence of anions is the dominant influence on the energy level structure. However, a key difference is observed at high anion concentrations in Figure 8, where an anion concentration of 0.66 is reached for 6-PEDOT +4. In this case, many energy levels cluster closely under the HOMO. This is caused by Coulombic repulsion between the anions, which causes their orbitals to rise in energy. This result can also be seen in [50] although it is not discussed. These orbitals are highly localized on the chlorine atoms of the anions, as shown in the HOMO to HOMO-2 of 6-PEDOT with four anions in Figure 9. Due to their high localisation, the anionic chlorine orbitals are able to have very similar energies to each other, hence the close stacking of their energy levels in Figure 8. This effect occurs further down the energy scale for 12-PEDOT with anions, because the overall anion concentration is only 0.33 instead of 0.66. This reduces the Coulombic repulsion between the anions and lowers the energy of their orbitals.

For 6-PEDOT with four anions, the HOMO-1 and HOMO-2 orbitals have dominant anionic character with a small amount of PEDOT character. The mixing of orbitals is due to a DFT error caused by the self-interaction error corrections. In redox systems including conducting polymers, the charge transfer is well known to be very close to 1, whereas DFT calculations predict 0.8-0.9. Therefore, it is reasonable to conclude that the HOMO-1 and HOMO-2 are pure anionic orbitals.

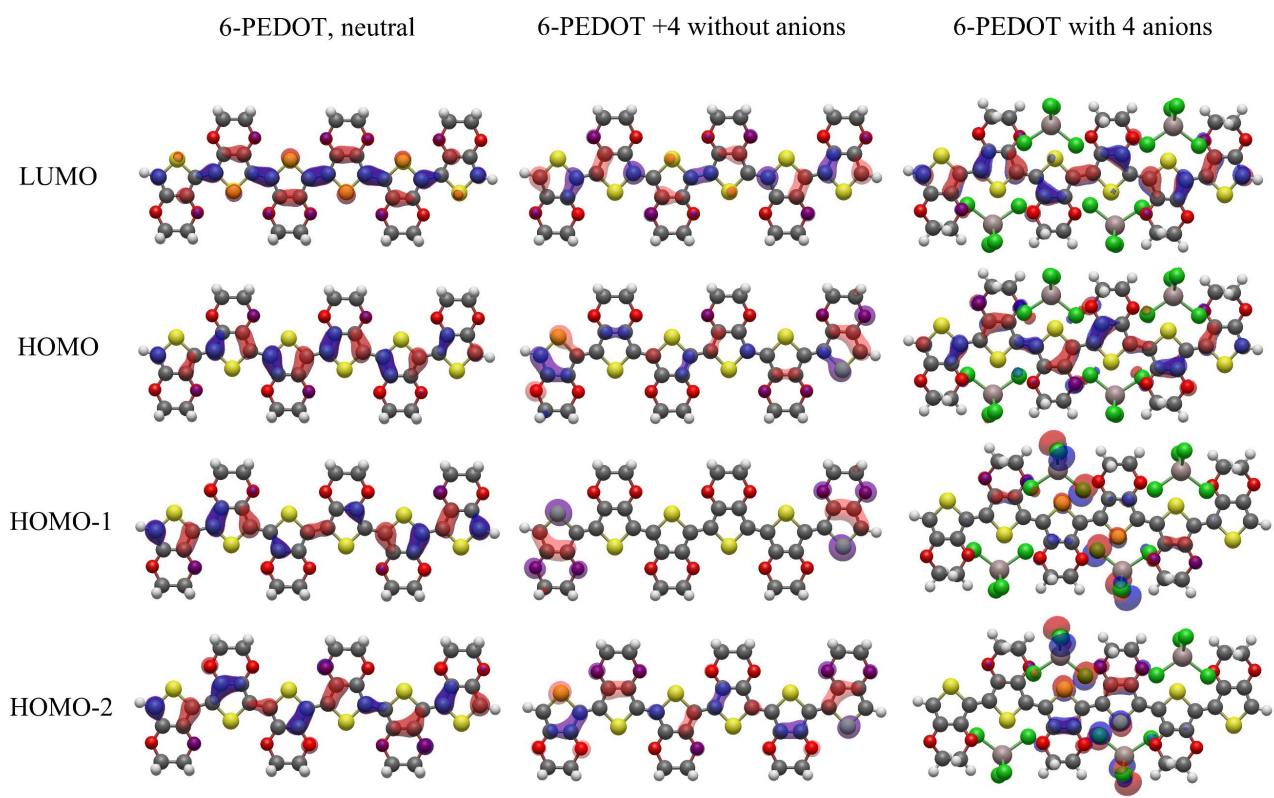

Figure 9 - Molecular orbitals, at isovalue 0.03, for neutral 6-PEDOT without counterions, 6-PEDOT with +4 oxidation state without counterions, and 6-PEDOT with four anions. The molecular orbitals show that the HOMO-1 and HOMO-2 for the four-anion case are highly localized on the chlorine atoms of the anions. These highly localized anionic orbitals mix with each other much less than the delocalized orbitals in the system without counterions and therefore are able to bunch more closely in energy than the orbitals in PEDOT without counterions. The HOMO-1 and HOMO-2 of the 6-PEDOT with four anions shows predominantly anionic orbitals with a contribution from the PEDOT orbitals: this is due to a DFT error, and the orbitals should be purely anionic.

## Effect of Coulombic repulsion between anions on conductivity

As discussed in the introduction, conductivity begins to decrease at high anion concentrations, after reaching a peak when about half the redox sites are occupied. Figure 9 shows that at anion concentrations of 0.66, localised anionic orbitals take the place of the HOMO-1 and HOMO-2, where at lower anionic concentrations these energy levels were PEDOT orbitals delocalised along the conjugated carbon backbone. The new localised anionic states are too far below the HOMO to interfere with conductivity at room temperature without nucleic motion – the HOMO-1 is around 0.1 eV below the HOMO (Figure 8). However, it is hypothesized that in a macropolymer with thermal nucleic motion, the natural variation in interatomic distances would cause these localised states to be regularly pushed up to a similar or greater energy than the HOMO, where the occupied and unoccupied redox sites that charge carriers hop between are located. This would interfere with the hopping process, lowering conductivity, and would affect both intra- and interchain conductivity. To the best of the authors' knowledge, this is a novel addition to the theory of conducting polymer conductivity.

## Study of the anion in isolation

In order to understand the nature of the states introduced by co-located anions, the anion was studied in isolation. $AlCl_4^-$ has a wide energy gap between its bonding and antibonding states (9.2 eV). When co-located with the PEDOT, the antibonding states sit at high energy, well above the HOMO-LUMO gap of the overall system as shown in Figure 8, while the bonding states sit at low energy, well below the gap. When the anion is considered in isolation, there are 8 p-derived non-bonding states on the $AlCl_4^-$ anion from HOMO to HOMO-7. The bonding, non-bonding and antibonding states are shown in Figure 9. All of the non-bonding states are chlorine p-orbitals of similar character to the anionic HOMO. It is clear from comparing Figure 9 and Figure 10 that the anionic orbitals appearing in the HOMO-1 and HOMO-2 for 6-PEDOT with four anions have the character of the anionic non-bonding orbitals. None of the anionic orbitals contribute to unoccupied energy levels appearing in the gap as the number of anions increases, in contrast to the predictions of the bipolaron model [29].

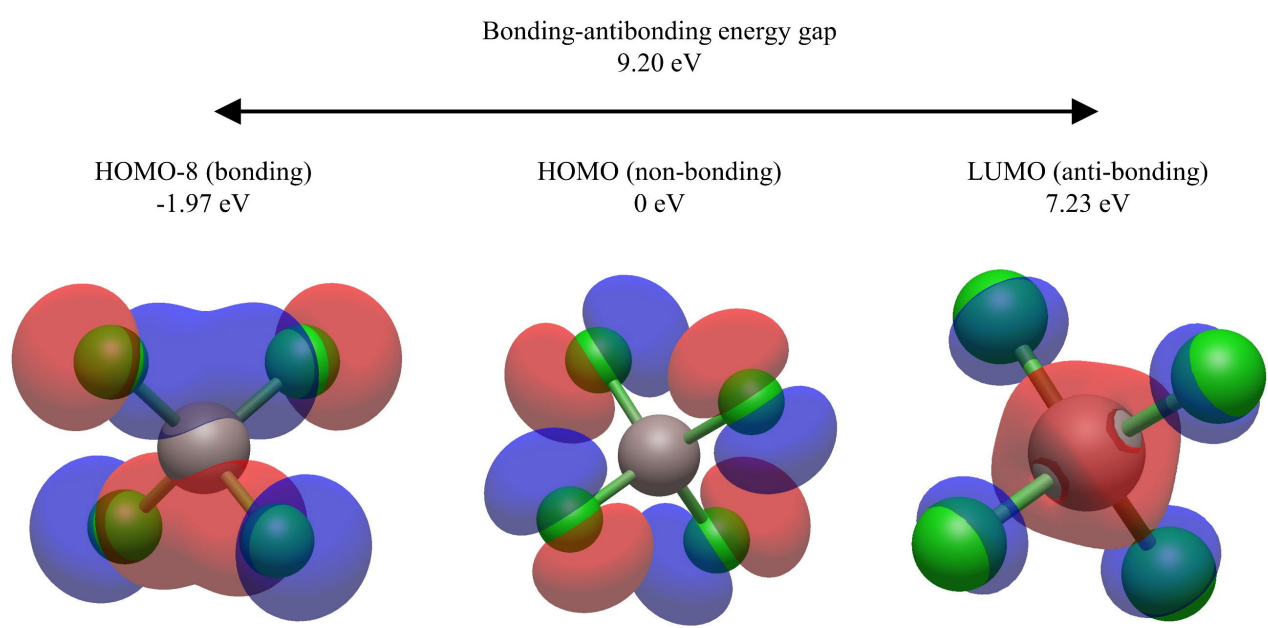

Figure 10 – Molecular orbitals at isovalue 0.03 for the HOMO-8 (bonding), HOMO (non-bonding) and LUMO (antibonding) of $AlCl_4^-$ with energy levels shown normalised to the HOMO of the isolated anion. The non-bonding orbitals are all equal mixtures of the chlorine p-orbitals of similar character to the HOMO, explaining why they sit very closely in energy in the case of 6-PEDOT with four anions in Figure 8.

## Conclusions

The results and analysis presented here add further evidence to move forwards from the original bipolaron model as developed from pre-DFT methods. In PEDOT with $AlCl_4^-$ anions, distortions localise around the anions, and when multiple anions are located near each other, a single longer distortion can appear across them. It has been found in this thesis and shown in the author's previously published work [15] that the initial anion locations are almost always stable, but that the anions do have Coulombic repulsion between them and will repel each other if concentration levels are too high. The lowest energy configurations have anions well distributed along the oligomer and not too close to the ends. This work leads naturally to a fuller exploration of the thermodynamics of the different anion configurations in Chapter 8.

Two distinct distortion characteristics have been observed in this work for PEDOT with and without counterions, identified by a single or a double peak in quinoid structure, as measured by carbon-carbon bond length inversions. It is noted that other numbers of distortion peaks, and different distortion shapes, are possible on longer chains with different anion configurations, with the bond length alternations correlated to the charge density in the PEDOT. The original bipolaron model is based on similar descriptions of distortions and uses the terms polarons, bipolarons or polaron pairs. However, the distortions observed in this work are not reliably associated with +1 or +2 states as predicted by any version of the bipolaron model. For PEDOT with anions, higher local anion concentrations are associated with double peak distortions appearing as the ground state. The double peak distortion may be similar to the distortion type observed experimentally for longer polythiophene oligomers without counterions with +2 oxidation state [47]. No reason is found to link distortion shapes to the terms polarons, bipolarons or polarons pairs. Instead, the distortions shapes are dependent on many aspects including oxidation state and configuration of anions, oligomer length, anion type and functional.

These findings add to the existing body of modern theoretical research disproving the original bipolaron model, often now referred to as the 'traditional models' in conducting polymer research. Some have chosen to adapt the terminology of the bipolaron model to reflect the modern understanding [6, 50]. Consequently, a more precise definition of polarons and bipolarons is suggested, that has already been partially developed by other authors. Here, the intent is to state the definition clearly and comprehensively. A polaron is a localised, singly charged state, caused in experiment by the introduction of an anion, and almost impossible to introduce in real life without an anion present. A polaron appears as a spin on an ESR measurement due to having a half-occupied HOMO. A bipolaron is what happens when two anions adsorb near enough to each other that the polarons overlap and the original HOMO becomes fully emptied local to the anions. This causes the ESR spins to disappear. This pattern repeats as increasing odd and even numbers of anions are introduced near to each other. There is no structural driver for anions to co-locate, and in fact they distribute evenly to minimise Coulombic repulsion. Consequently, the main difference between polarons and bipolarons is that polarons are a localized state having spin asymmetry while bipolarons are a localised state having spin symmetry. Apart from that, there are no further differences between polarons and bipolarons, and no definite limit on the number of anions that can share a single distortion.

The even distribution of anions throughout the polymer to begin with means that ESR spins build up during charging, similar to the concept of the polaron pair as proposed by Sahalianov *et al.*, though as discussed this has not been fully explored in this thesis [50]. Then as the anion concentration rises the anions are forced to locate nearer to each other, so that the polarons overlap and the half-occupied levels become unoccupied and the spins disappear. While each anion introduces one bonding and one antibonding state, these are far below and above the HOMO-LUMO gap respectively. The apparent new state appearing in the HOMO-LUMO gap is the newly unoccupied half or whole delocalised PEDOT energy level which has been pushed up by the occupied levels below it into the gap, with the non-bonding anionic orbitals clustering below the HOMO.

At the highest anion concentration modelled of 0.66, the non-bonding orbitals of the anions are promoted to an energy level approaching that of the system HOMO, driven by

Coulombic repulsion between the anions. It is hypothesized that in the macropolymer, these highly localised states would be regularly promoted to the energy of the sites responsible for charge carrier hopping as a result of nucleic motion, interfering with both intra- and interchain conductivity. The proposed interference with the hopping mechanism(s) is hypothesized to contribute to, or cause, the loss of conductivity found in conducting polymers at anion concentrations above 0.5. This hypothesis is a novel contribution to conducting polymer theory.

## Supplementary Material

The energy level structure in Figure 8 was based on 6- and 12-PEDOT with anions. The following table shows where the anions are located on the oligomers. These are the lowest energy states found for that number of anions from either a complete sample for 6-PEDOT or large random sample for 12-PEDOT.

| Number of anions | 6-PEDOT | 12-PEDOT |
|---|---|---|
| 1 | 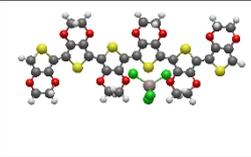 | 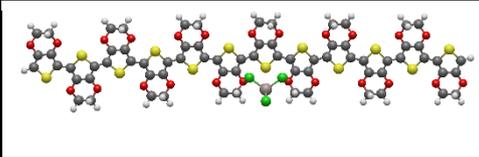 |
| 2 | 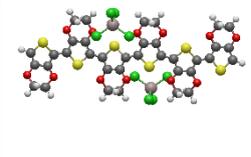 | 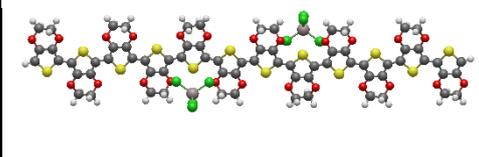 |
| 3 | 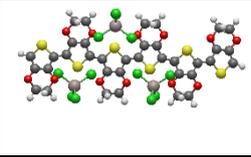 | 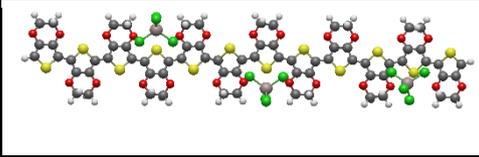 |
| 4 | 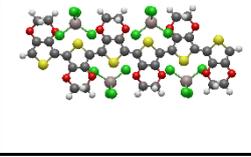 | 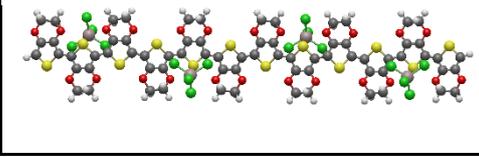 |

Figure 11 – the relaxed configurations of the PEDOT oligomers with anions for which the energy levels are presented in Figure 8. Anions are placed either above or below the oligomer, which can be observed via overlaps in the molecular drawings.

## Acknowledgements

The authors acknowledge support from the International Consortium of Nanotechnologies (ICON) funded by Lloyd's Register Foundation [G0086], a charitable foundation which helps to protect life and property by supporting engineering-related education, public engagement and the application of research, and from the Engineering and Physical Sciences Research Council, through the Centre for Doctoral Training in Energy Storage and its Applications at the University of Southampton [EP/L016818/1]. We are grateful to the UK Materials and Molecular Modelling Hub for computational resources, which is partially funded by EPSRC (EP/T022213/1, EP/W032260/1 and EP/P020194/1), and for the use of the

IRIDIS High Performance Computing Facility, and associated support services at the University of Southampton.

## Conflict of Interest Statement

The authors have no conflicts to disclose.

## Author Contributions Statement

**Ben Craig**: Conceptualization (lead), Writing/Original Draft Preparation (lead), Writing/Review and Editing (equal). **Peter Townsend**: Writing/Review and Editing (equal). **Denis Kramer**: Writing/Review and Editing (equal), Supervision (equal). **Chris-Kriton Skylaris** and **Carlos Ponce de Leon**: Supervision (equal).

## Data availability

The data that support the findings of this study are available from the corresponding author upon reasonable request.